\documentclass[twocolumn,showpacs,preprintnumbers,amsmath,amssymb]{revtex4}

\usepackage{color}
\usepackage{graphics}% Include figure files
\usepackage{dcolumn}% Align table columns on decimal point
\usepackage{bm}% bold math

\renewcommand\textfraction 0
\renewcommand\topfraction 1
\renewcommand\bottomfraction 1
\begin{document}
%\begin{frontmatter}
%\title{{\normalsize Submitted to J. Supercond:INM}\\
%\vspace*{.8cm}
\title{
Magnetic phase diagram of Ca$_{1-x}$Mn$_x$O} \author{S. Kolesnik and B. Dabrowski}
\affiliation{Department of Physics, Northern Illinois University, DeKalb, IL 60115}
\date{\today}
\begin{abstract}
Alternating current susceptibility and direct current magnetization have been studied for
polycrystalline Ca$_{1-x}$Mn$_x$O. On increasing the Mn content, magnetic ordering changes from
spin-glass behavior for $0.25 \leqslant x \leqslant 0.4$ to antiferromagnetic order. The
paramagnetic/antiferromagnetic transition is of second order for $0.5 \leqslant x \leqslant 0.65$
and of first order for $x \geqslant 0.7$. For low Mn concentrations, the high-temperature
alternating current susceptibility can be described by a diluted Heisenberg magnet model developed
for diluted magnetic semiconductors.

%\vspace*{.5cm} Keywords: Diluted magnetic semiconductor, Zn$_{1-x}$Mn$_x$O, synthesis, spin-glass
%behavior\\
\end{abstract}

%\begin{keyword}
%\end{keyword}

%}
\maketitle
%\end{frontmatter}

 Theoretical predictions of
room temperature ferromagnetism in Mn-containing diluted magnetic semiconductors \cite{Dietl00}
recently brought wide attention to this class of materials. According to these calculations,
$p$-type Zn$_{1-x}$Mn$_x$O is a promising candidate for a room temperature ferromagnet.
Pulsed-laser deposited Zn$_{1-x}$Mn$_x$O thin films without intentional carrier doping show
spin-glass behavior.\cite{Fukumura01} According to this latter study, Mn can be dissolved in the
ZnO matrix to over 35\%. We have shown in a previous paper that the solubility of Mn in the
zinc-blende structure of ZnO is less than 15\% for polycrystalline samples under various oxygen
pressure conditions. Samples obtained in air or argon are paramagnetic, while the high-pressure
oxygen annealing induces spin-glass-like behavior by precipitation of ZnMnO$_3$ in the paramagnetic
matrix.\cite{Kolesnik02}

In this study we investigate polycrystalline Ca$_{1-x}$Mn$_x$O $(x = 0.125-1)$. Both end-member
compounds, CaO and MnO, adopt the rock-salt structure and the complete range of solid solution can
be achieved. This makes it possible to study a continuous evolution of magnetic properties from
diamagnetic insulating/semiconducting CaO (energy gap $E_g = 7.09$~eV)\cite{Whited69} to
antiferromagnetic insulating MnO ($T_N \simeq 120$~K, $E_g = 4$~eV), whose antiferromagnetic
properties were already observed seventy years ago.\cite{Tyler33} Due to similarity of the ionic
radii (0.8~\AA~for Mn$^{2+}$ and 0.99~\AA~for Ca$^{2+}$) Mn$^{2+}$ ions should occupy a
substitutional site in CaO in contrast to BaO (ionic radius of Ba$^{2+}$ is equal to 1.34~\AA),
where Mn$^{2+}$ ions locate off-center.\cite{Winsum78} The CaO-MnO solid solution has been shown to
form in the H$_2$/He atmosphere above 1173 K from carbonate precursors.\cite{Poeppelmeier86}
Thermodynamic functions of mixing have been determined at high temperatures for
Ca$_{1-x}$Mn$_x$O.\cite{Rog92} To our knowledge, the magnetic properties of this solid solution
have not yet been reported. MnO undergoes a first-order transition to the antiferromagnetic state
at $T_N = 115-120$~K \cite{Seino73, Srinivasan83} This transition is accompanied by a trigonal
deformation\cite{Morosin70} to a slightly rhombohedral structure, compressed along the [111]
direction.\cite{Bloch73} The antiferromagnetic order, determined by neutron diffraction, is of
type-II (AFII) in the fcc lattice and consists of (111) sheets of ferromagnetically ordered
Mn$^{2+}$ ions, which are antiferromagnetically coupled to the neighboring sheets along the [111]
direction.\cite{Roth58} Magnetic properties of MnO have been explained in terms of two
superexchange constants $J_1 = 10$~K and $J_2 = 11$~K for the nearest-neighbor (NN) and
next-nearest-neighbor (NNN) interactions, respectively.\cite{Lines65} Recent local-density
approximation (LDA) calculations\cite{Pask01} suggest relatively stronger NNN interactions $J_1 =
9.8$~K and $J_2 = 24.5$~K. The small uniaxial stress along the [111] axis of the MnO crystal
results in the clearly first-order transition and the increasing $T_N$.\cite{Bloch73} This result
has been ascribed to the strain-related suppression of T domains (which correspond to
antiferromagnetic stacking along the four [111] directions within the crystal). Due to the presence
of these T domains the first-order character of the antiferromagnetic/paramagnetic transition may
be obscured. Recent heat capacity studies showed a continuous magnetic transition in small,
stress-free, MnO crystals.\cite{Woodfield99} The application of large [111] stress ($1.2 < \tau
\leqslant 5.5$~kbar) leads to a continuous transition, again. \cite{Bloch75} Solid solution
Mn$_{1-x}$Ni$_x$O shows a linear increase of $T_N$ upon increasing Ni concentration with
preservation of the AFII structure.\cite{Cheetham83} Both end members of this series, MnO and NiO
($T_N \simeq 525$~K) show the same type of rhombohedral crystallographic distortions.\cite{Roth58}

The mixture of CaCO$_3$ and MnO$_2$ was calcined in air several times at $T = 900, 1000,
1100^{\circ}$C with intermediate grindings. The final synthesis stage was performed in a hydrogen
atmosphere at $T = 1050^{\circ}$C. The alternating current (ac) susceptibility and direct current
(dc) magnetization were measured using a Physical Property Measurement System (Quantum Design). The
ac susceptibility was measured in the range 2.5-395~K, in the excitation field $H_{ac} = 14$~Oe at
the frequency $\omega = 1$~kHz. The ``zero-field-cooled'' ($M_{ZFC}$) and ``field-cooled''
($M_{FC}$) magnetizations were measured in a magnetic field of 1 kOe. $M_{ZFC}$ was measured on
warming after cooling in a zero magnetic field and switching the magnetic field on at $T = 5$~K.
$M_{FC}$ was subsequently measured on cooling in the magnetic field.

\begin{figure}[!]
\resizebox{8.5cm}{!}{\includegraphics{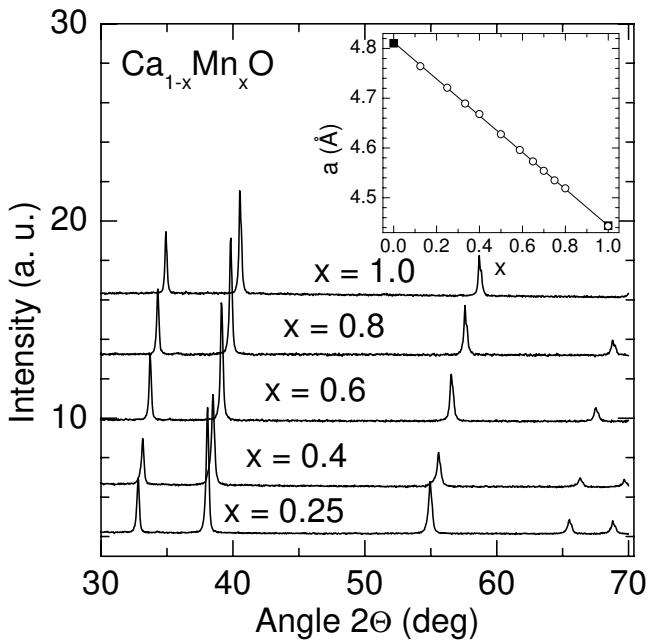}}
 \caption{\label{xray} X-ray
diffraction patterns for several Ca$_{1-x}$Mn$_{x}$O samples. Inset: lattice parameter, $a$, of the
rock-salt structure of Ca$_{1-x}$Mn$_{x}$O. The solid line is a linear fit to the data. Solid
squares are literature values for CaO and MnO. }
\end{figure}

X-ray diffraction spectra have been collected using a Rigaku X-ray diffractometer. The sharp X-ray
diffraction patterns, presented in Fig.~\ref{xray} for several samples, indicate good quality,
homogeneous material.
The single-phase rock-salt-type structure is revealed for all the Ca$_{1-x}$Mn$_x$O samples.  Inset
to Fig.~\ref{xray} shows the lattice parameter, $a$, determined from the X-ray diffraction patterns
using the GSAS refinement software. The lattice parameter, $a$,  is linearly dependent on the Mn
content $x$ with a very good accuracy. The linear fit to the experimental data gives a dependence
$a(x) = 4.812(1) - 0.369(2)x$~(\AA). For CaO, the extrapolated $a$ is equal to 4.812(1)~\AA, which
is in agreement with the literature value $a = 4.81$~\AA.\cite{McMurdie86} For our MnO sample, $a =
4.443(3)$~\AA~is very close to the literature value $a = 4.442-4.446$~\AA.\cite{Barrett64} A very
good agreement between the measured lattice parameters and the linear dependence for all Mn
concentrations suggests that the actual compositions and the nominal ones are the same.

Three different kinds of magnetic transitions observed for Ca$_{1-x}$Mn$_x$O are illustrated in
\begin{figure*}[t]
\resizebox{16cm}{!}{\includegraphics{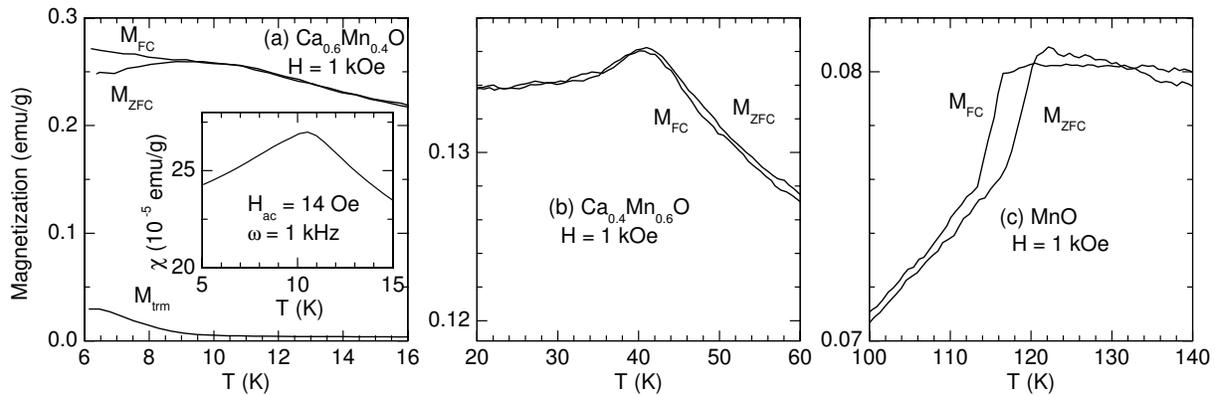}} \caption{\label{zfr} ``Zero-field-cooled''
($M_{ZFC}$), ``field-cooled'' ($M_{FC}$), and thermoremanent ($M_{trm}$) magnetizations  for
several Ca$_{1-x}$Mn$_x$O samples.  The inset shows the cusp in ac susceptibility. No
thermoremanent magnetization was observed for Ca$_{0.4}$Mn$_{0.6}$O or MnO. }
\end{figure*}
Fig.~\ref{zfr}. For $x = 0.25-0.4$, the spin-glass behavior is observed at low temperatures. A
difference between the ``zero-field-cooled'' ($M_{ZFC}$) and ``field-cooled'' ($M_{FC}$)
magnetizations, which is characteristic of spin-glass behavior, is shown in Fig.~\ref{zfr}(a) for
$x = 0.4$. The presence of the thermoremanent ($M_{trm}$) magnetization at low temperatures
supports this observation. The inset to Fig.~\ref{zfr}(a) shows a cusp in the ac susceptibility.
The temperature, at which the cusp is maximum defines the spin-glass freezing temperature, $T_f$.
Similar cusp was observed for the $x = 0.333$ sample and a kink on the ac susceptibility for $x =
0.25$. The same type of spin-glass behavior has been previously observed for Mn-containing diluted
magnetic semiconductors (e.g. Zn$_{1-x}$Mn$_x$Se~\cite{Twardowski87} and
Cd$_{1-x}$Mn$_x$Te~\cite{Denissen87}), where a cusp was observed for high Mn concentrations and a
kink for low Mn concentrations.

For $x = 0.5-0.65$, $M_{ZFC}$ and $M_{FC}$ curves trace each other and no thermoremanent
magnetization is observed. This behavior, presented in  Fig.~\ref{zfr}(b) indicates  a continuous
(second-order) paramagnetic/antiferromagnetic transition. For higher Mn concentrations, $x
\geqslant 0.7$ including MnO, the paramagnetic/antiferromagnetic transition is accompanied by a
hysteresis, clearly observed both in the ac susceptibility and dc magnetization (see:
Fig.~\ref{zfr}(c)). This hysteresis is a sign of the discontinuous first-order transition.

\begin{figure}[!]
\resizebox{8.5cm}{!}{\includegraphics{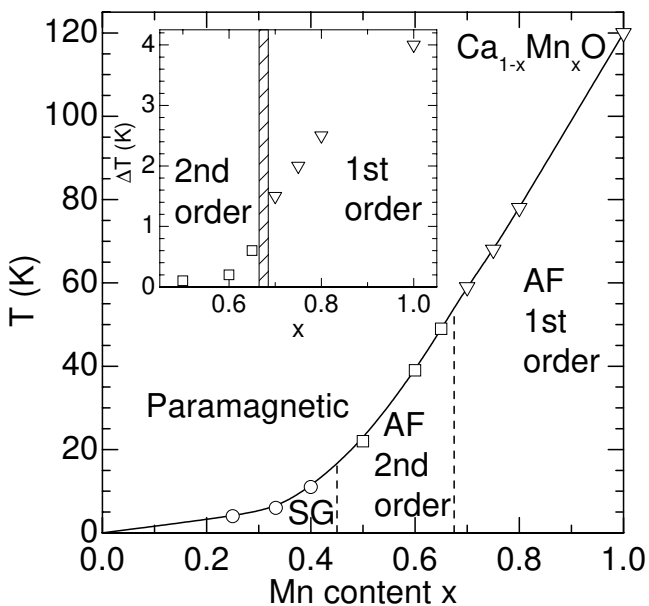}} \caption{\label{phd} Magnetic phase diagram of
Ca$_{1-x}$Mn$_x$O. SG denotes spin-glass behavior. Two antiferromagnetic phases to which the
transition is of first or second order are denoted as ``AF 1st order'' and ``AF 2nd order'',
respectively. The solid line is a guide to the eye. The inset shows the width of the thermal
hysteresis in the ac susceptibility at the AF transition.}
\end{figure}

The observed magnetic transition temperatures are collected in  Fig.~\ref{phd}.
The $x = 0.125$ sample is paramagnetic down to $T = 2.5$~K. The $x = 0.25-0.4$ samples show
spin-glass behavior at increasing temperatures with increasing Mn content. The antiferromagnetic
transition temperatures for $x \geqslant 0.5$ are defined as the temperatures for which the
temperature dependence of the ac susceptibility has the maximum slope, i.e. slightly below the
observed maximum of $\chi(T)$. For the first-order transitions, ($x \geqslant 0.7$), we used the
transition temperatures, which were determined on warming ($M_{ZFC}$). The transition temperatures
in all the Mn concentration regions seem to continuously change with $x$, but no common scaling has
been determined. The spin-glass behavior at low Mn concentrations is quantitatively similar to the
behavior of wurtzite and zinc-blende diluted magnetic semiconductors. The temperature dependence of
the ac susceptibility for $x \geqslant 0.5$ is almost identical with that of the dc magnetization.
The width of the thermal hysteresis $\Delta T$ in the ac susceptibility at the antiferromagnetic
transition, which is shown in the inset to Fig.~\ref{phd} starts increasing with Mn concentration
for $x \approx 2/3$. This indicates a crossover between the second-order and first-order nature of
the transition.

 The high-temperature ac susceptibility of Ca$_{1-x}$Mn$_x$O generally shows Curie-Weiss behavior.
Within the framework of the diluted Heisenberg antiferromagnet theory,\cite{Spalek86} the
high-temperature ac mass susceptibility can be described by the formula
\begin{equation}
 \chi = \frac{C_M(x)}{\mu(x)(T-\Theta(x))},
\end{equation}
where $\chi$ is the ac susceptibility after subtraction of the diamagnetic contribution of the host
material CaO (equal to $-0.27 \cdot 10^{-6}$ emu/g) and  $C_M(x)$ is the molar Curie constant
defined as
\begin{equation}
 C_M(x) = x\frac{N(g\mu_B)^2S(S+1)\mu(x)}{3k_B\rho(x)},
\end{equation}
$N$ is the number of cations per unit volume, $S$ is the effective spin of Mn$^{2+}$ ion, $g = 2$
is the gyromagnetic factor, $\mu(x)$ is the molar mass, $\rho(x)$ is the mass density, $\Theta(x) =
\Theta_0 \cdot x$ is the Curie-Weiss temperature
\begin{equation}
\Theta_0(x) = -\frac{2}{3}xS(S+1)\sum_P J_Pz_P/k_B\equiv \Theta_0x,
\end{equation}
where $J_P$ is the exchange integral for a pair of $P$th Mn neighbors and $z_P$ is the number of
neighbors of $P$th order. In the diluted magnetic regime the dominating interactions are between
the nearest neighbors. Hence, we may limit the expansion in Eq. (3) to the first order.
 The $\Theta_0$ constant is then related to the exchange integral between the
nearest Mn neighbors $J_1$,
\begin{equation}
\frac{2J_1}{k_B} = \frac{3\Theta_0}{zS(S+1)},
\end{equation}
where $z = 12$ is the number of nearest neighbors in the rock-salt structure of Ca$_{1-x}$Mn$_x$O.

\begin{figure}[!]
\resizebox{8.5cm}{!}{\includegraphics{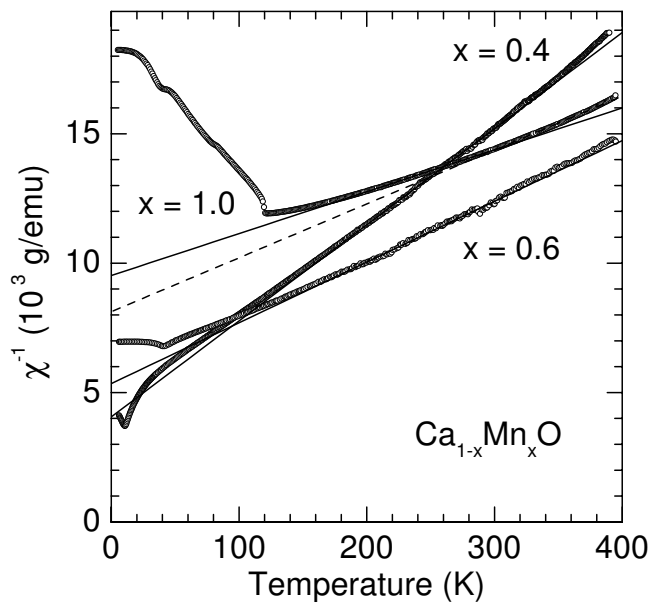}}
 \caption{\label{invchi} Inverse ac susceptibility for Ca$_{1-x}$Mn$_x$O samples. The solid lines are linear fits of to the
data in the range 200-300~K. The dashed line is a linear fit for $x = 1$ (i.e. MnO) in the range
300-390~K.}
\end{figure}

The inverse ac susceptibility for several Ca$_{1-x}$Mn$_x$O samples is shown in Fig.~\ref{invchi}.
For low Mn concentrations, $x \leqslant 0.333$, the inverse ac susceptibility is linear with
temperature in a wide range $T = 50-390$~K. Deviations from linearity (characteristic to diluted
antiferromagnets and related to the interactions between next nearest neighboring Mn$^{2+}$
ions\cite{Spalek86}) are observed at lower temperatures. For higher Mn concentrations, the slope of
the inverse susceptibility changes with temperature. We have analyzed the linear fits to the
$\chi^{-1}(T)$ data in three temperature ranges: 100-200~K, 200-300~K, and 300-390~K.

\begin{figure}[!]
\resizebox{8.5cm}{!}{\includegraphics{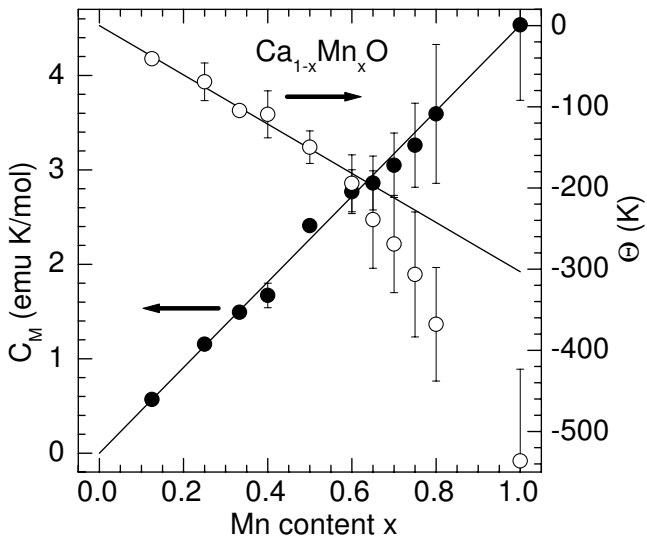}}
 \caption{\label{cw} Molar Curie constant $C_M(x)$ (solid circles) and Curie-Weiss temperatures $\Theta(x)$
(open circles) for Ca$_{1-x}$Mn$_x$O samples. The solid lines are linear fits to the first three
data points.}
\end{figure}

The data points in Fig.~\ref{cw} represent the parameters determined from the fit of Eq. (1) to the
inverse susceptibility data in the temperature range 200-300~K.
The values determined from the fits in the other two temperature regions are plotted as error bars.
For a given $x$, the fit in the lower temperature range (100-200 K) gives higher absolute
$\Theta(x)$ and larger $C_M(x)$ than the fit in the higher temperature range (300-390 K).

 For diluted Heisenberg
antiferromagnets it is expected that the material parameters $C_M(x)$ and $\Theta(x)$ are
proportional to $x$. We have made linear fits to the first three data points (i.e. $0.125 \leqslant
x \leqslant 0.333$, assuming that $C_M(x=0) = 0$ and $\Theta(x=0) = 0$). From these fits, we have
obtained the extrapolated values $C_M(x=1) = 4.53(4)$~emu K/mol and $\Theta_0 = -303(11)$~K. The
value of $C_M$ is close to those determined for Mn-containing II-VI diluted magnetic
semiconductors.\cite{Spalek86} From Eq. (2), calculated $S = 2.55(5)$, which is in perfect
agreement with the effective spin of Mn$^{2+}$ ion $S = 5/2$. The extrapolated $\Theta_0$ is lower
(in the sense of the absolute value) than the values found for diluted magnetic semiconductors
($\Theta_0 = -470$ to $-831$~K).\cite{Spalek86} We observe significant deviations from the linear
behavior of $\Theta(x)$ for $x \geqslant 0.5$, where the antiferromagnetic order can be observed
and the diluted magnet model is no longer valid. It is worth noting that for higher Mn
concentrations the Curie-Weiss temperatures become comparable to or higher than our experimental
temperature range, which increases the error of the determined values. However, we suggest that
these deviations of $\Theta(x)$ from the linear behavior can be explained by the increasing
contribution of NNN interactions between highly concentrated Mn$^{2+}$ ions.

By including the NNN interactions we obtain the following modification of Eq. (3)
\begin{equation}
z\frac{2J_1}{k_B} + z_2\frac{2J_2}{k_B} = \frac{3\Theta(x)}{S(S+1)},
\end{equation}
where $z_2 = 6$ is the number of next-nearest neighbors. For MnO, for example, $\Theta = -536$~K.
By using $2J_1/k_B = -8.7(3)$~K calculated from the extrapolated $\Theta_0 = -303$~K according to
Eq. (4), we obtain $2J_2/k_B \approx -13.2$~K for MnO, which gives us a qualitative agreement with
the LDA calculations\cite{Pask01}, namely $J_2
> J_1$. However, the actual values of our exchange constants are significantly smaller than those found in the
literature (see: references to Ref. \cite{Pask01}). Note that M. E. Lines and E. D.
Jones\cite{Lines65} used an analogical formula to our Eq. (5) but without the factor 2. From their
$\Theta = -540$~K, this gave as a result  $(2J_2+J_1)/k_B= 30.9$~K, which is twice as higher as our
value.

 In summary, we have determined the magnetic phase diagram of Mn- doped polycrystalline CaO samples. On
increasing the Mn content, magnetic ordering changes from the spin-glass behavior for $0.25
\leqslant x \leqslant 0.4$ to the antiferromagnetic order. The paramagnetic/antiferromagnetic
transition is second order for $0.5 \leqslant x \leqslant 0.65$ and first order for $x \geqslant
0.7$. For low Mn contents, the ac susceptibility can be described by a diluted Heisenberg magnet
model developed for diluted magnetic semiconductors.

This work was supported by the DARPA/ONR and the State of Illinois under HECA.


\begin{thebibliography}{99}

\bibitem{Dietl00} T. Dietl, H. Ohno, F. Matsukura, J. Cibert, and D. Ferrand, Science {\bf 287}, 1019 (2000).
\bibitem{Fukumura01} T. Fukumura, Z. Jin, M. Kawasaki, T. Shono, T. Hasegawa, S. Koshihara, and H. Koinuma, Appl. Phys. Lett. {\bf 78}, 958 (2001).
\bibitem{Kolesnik02} S. Kolesnik, B. Dabrowski, and J. Mais, J. Supercond.: Incorp. Novel
Magnetism {\bf 15}, 251 (2002).
\bibitem{Whited69} R. C. Whited and W. C. Walker, Phys. Rev. {\bf 188}, 1380 (1969).
\bibitem{Tyler33} R. W. Tyler, Phys. Rev. {\bf 44}, 776 (1933).
\bibitem{Winsum78} J. A. van Winsum, T. Lee, and H. W. den Hartog, Phys. Rev. B {\bf 18}, 173 (1978).
\bibitem{Poeppelmeier86} K. R. Poeppelmeier. H. S. Horowitz, and J. M. Longo, J. Less-Common Metals
{\bf 116}, 219 (1986).
\bibitem{Rog92} G. R\'og, A. Koz{\l}owska-R\'og, W. Pycior, and K. Zaku{\l}a, J. Solid State Chem.
{\bf 100}, 115 (1992).
\bibitem{Seino73} D. Seino, S. Miyahara, and Y. Noro, Phys. Lett. {\bf 44A}, 35, (1973).
\bibitem{Srinivasan83} G. Srinivasan and M. S. Seehra, Phys. Rev. B {\bf 28}, 6542, (1983).
\bibitem{Morosin70} B. Morosin, Phys. Rev. B {\bf 1}, 236, (1970).
\bibitem{Bloch73} D. Bloch and R. Maury, Phys. Rev. B {\bf 7}, 4883 (1973).
\bibitem{Roth58} C. G. Shull, W. A. Strausser, and E. O. Wollan, Phys. Rev. {\bf 83}, 333 (1951);
W. L. Roth, Phys. Rev. {\bf 110}, 1333 (1958).
\bibitem{Lines65} M. E. Lines and E. D. Jones, Phys. Rev. {\bf 139}, A1313 (1965).
\bibitem{Pask01} J. E. Pask, D. J. Singh, I. I. Mazin, C. S. Hellberg, and J. Kortus, Phys. Rev. B {\bf 64}, 024403 (2001).
\bibitem{Woodfield99} B. F. Woodfield, J. L. Shapiro, R. Stevens, J. Boerio-Goates, and M. L. Wilson, Phys. Rev. B {\bf
60}, 7335, (1999).
\bibitem{Bloch75} D. Bloch, D. Hermann-Ronzaud, C. Vettier, W. B. Yelon, and R. Alben, Phys. Rev. Lett. {\bf 35}, 963 (1975).
\bibitem{Cheetham83} A. K. Cheetham and D. A. O. Hope, Phys. Rev. B {\bf 27}, 6964 (1983).
\bibitem{McMurdie86} H. McMurdie {\em et al.}, Powder Diffraction {\bf 1}, 266 (1986); B. Reardon and C. Hubbard, TM-11948, Oak Ridge Natl. Lab. Rep. ORNL (U.S.),
(1992).
\bibitem{Barrett64} C. A. Barrett and E. B. Evans, J. Am. Ceram. Soc. {\bf 47}, 533 (1964);
S. Sasaki {\em et al.}, Acta Crystallogr., Sec. A {\bf 36}, 904 (1980).
\bibitem{Twardowski87} A. Twardowski, H. J. M. Swagten, W. J. M. de Jonge, and M. Demianiuk, Phys. Rev. B {\bf 36}, 7013 (1987).
\bibitem{Denissen87} C. J. M. Denissen, S. Dakun, K. Kopinga, W. J. M. de Jonge, H. Nishihara, T. Sakakibara, and T. Goto, Phys. Rev. B {\bf 36}, 5316 (1987); R. R.
Ga{\l}\c{a}zka, S. Nagata, and  P. H. Keesom, Phys. Rev. B {\bf 22}, 3344 (1980)
\bibitem{Spalek86} J. Spa{\l}ek, A. Lewicki, Z. Tarnawski, J. K. Furdyna, R. R. Ga{\l}\c{a}zka, and Z. Obuszko, Phys. Rev. B {\bf 33}, 3407 (1986).
\end{thebibliography}
\end{document}